\pdfoutput=1
\documentclass[11pt,a4paper]{article}

\usepackage[utf8]{inputenc}
\usepackage[T1]{fontenc}
\usepackage{lmodern}
\usepackage[margin=1in]{geometry}
\usepackage{amsmath,amssymb,amsthm}
\usepackage{graphicx}
\usepackage{booktabs}
\usepackage{xcolor}
\usepackage{natbib}
\usepackage{hyperref}
\usepackage{cleveref}
\usepackage{enumitem}
\usepackage{caption}
\usepackage{multirow}
\usepackage{makecell}
\usepackage{mdframed}
\usepackage{algorithm}
\usepackage{algpseudocode}

\hypersetup{
  colorlinks=true,
  linkcolor=blue!70!black,
  citecolor=green!50!black,
  urlcolor=blue!60!black,
}

\newcommand{\tpw}{\text{tok/W}}

\newcommand{\nmax}{n_{\max}}
\newcommand{\nact}{n_{\text{active}}}
\newcommand{\Pidle}{P_{\text{idle}}}
\newcommand{\Pnom}{P_{\text{nom}}}
\newcommand{\Prange}{P_{\text{range}}}

\title{%
  The $1/\mathcal{W}$ Law: An Analytical Study of\\
  Context-Length Routing Topology and GPU Generation Gains\\
  for LLM Inference Energy Efficiency%
}

\author{%
  Huamin Chen$^{1}$ \quad
  Xunzhuo Liu$^{1}$ \quad
  Yuhan Liu$^{2}$ \\[4pt]
  Junchen Jiang$^{3}$ \quad
  Bowei He$^{4}$\thanks{Corresponding author: \texttt{Bowei.He@mbzuai.ac.ae}} \quad
  Xue Liu$^{4}$
  \\[6pt]
  $^{1}$vLLM Semantic Router Project \\[1pt]
  $^{2}$University of Chicago \quad
  $^{3}$Tensormesh Inc / University of Chicago \\[1pt]
  $^{4}$MBZUAI / McGill University
}
\date{2026}

\begin{document}
\maketitle

\begin{abstract}
How many tokens can a GPU inference cluster deliver per watt?  Across
deployments of identical hardware, the answer varies by $40\times$---not
because of software inefficiency, but because of the serving context window.
We derive the $1/\mathcal{W}$ \emph{law}: tokens per watt halves every time
the context window doubles.  A larger context window shrinks the KV-cache
concurrency limit while leaving GPU power draw roughly unchanged.  At 64K
context, an H100 holds 16 sequences in flight (tok/W $= 1.5$); at 4K context,
the same H100 holds 256 sequences (tok/W $= 17.6$).

Routing topology---which determines the effective context window each GPU
services---is a more powerful energy lever than buying newer hardware.
Working from published H100 power measurements~\citep{chung2026mlenergy}, a
calibrated logistic power model~\citep{liang2026g2g}, and a roofline throughput
model~\citep{xu2026aiconfigurator}, we derive these results analytically using
the inference-fleet-sim framework~\citep{chen2026fleetsim}; no new hardware
experiments were conducted.  Two-pool context-length routing
(FleetOpt~\citep{chen2026fleetopt}) delivers roughly $2.5\times$ better tok/W
over a homogeneous fleet, while upgrading from H100 to B200 delivers roughly
$1.7\times$.  The gains are independent: combining FleetOpt with B200 yields
$4.25\times$ over the H100 homogeneous baseline.  B200/H200 numbers are
analytical projections ($\pm$20\% uncertainty); H100 results are calibrated
to published measurements.

For MoE models, active-parameter weight streaming adds a third lever.
Qwen3-235B-A22B (22B active) reaches roughly 37.8 tok/W at 8K context on
H100---$5.1\times$ better than Llama-3.1-70B---because decode time scales with
activated weights, not total parameters.  MoE dispatch overhead is excluded,
so this is an upper bound.
\end{abstract}

% ────────────────────────────────────────────────────────────────────────────
\section{Introduction}
\label{sec:intro}

Generative AI workloads now consume a measurable fraction of data center
power, and that fraction is growing.  Unlike batch compute jobs, LLM inference
has a nonlinear relationship between load and power: a GPU serving a single
request idles near 43\% of its thermal design power (TDP), while the same GPU
running a large batch approaches 86\% TDP~\citep{chung2026mlenergy}.  The
number of requests processed in parallel is therefore not just a throughput
concern---it directly determines how much useful work you get per joule.

The main thing controlling how many requests a GPU can hold in parallel is the
serving context window.  A KV cache of fixed VRAM can hold fewer concurrent
sequences as the per-token memory footprint grows.  A Llama-3.1-70B deployment
on H100 (80~GB, TP=8 with tensor-parallel KV sharding)~\citep{meta2024llama3}
can hold 256 concurrent 4K-token sequences but only 16 concurrent
64K-token sequences.  Tokens per
watt drops by the same factor across those two operating points---roughly
$12\times$---and extends to a nearly $40\times$ spread across the full
2K to 128K context range.

GPU vendors have been responding to growing context demands with more
powerful accelerators.  B200 replaces H100 with 2.4$\times$ the memory bandwidth and 2.4$\times$
the VRAM---but also a TDP that is 43\% higher (1{,}000~W vs.\ 700~W).  Whether you actually get better tokens per watt
from a B200 depends entirely on the context window distribution your workload
presents at inference time.

This paper works out the quantitative relationships between context window,
routing topology, hardware generation, and fleet energy efficiency.  We connect three existing tools---the logistic GPU power model of Liang et
al.~\citep{liang2026g2g} calibrated to ML.ENERGY measurements~\citep{chung2026mlenergy},
the AIConfigurator roofline~\citep{xu2026aiconfigurator},
and the \texttt{inference-fleet-sim} planner~\citep{chen2026fleetsim}---to derive
the $1/\mathcal{W}$ law (Section~\ref{sec:ctx-lever}): tok/W halves every time
the serving context window doubles, because halving concurrency halves throughput
while power stays flat.  This makes the context window---not the GPU---the
primary energy knob.

From there we show (Section~\ref{sec:topo-effects}) that topology and hardware
generation are independent levers.  FleetOpt routing gives roughly $2.5\times$
over homogeneous on both H100 and B200; B200 gives roughly $1.7\times$ over H100
at any topology.  Because the gains are orthogonal they multiply, reaching
$4.25\times$ combined.  Neither lever alone gets you half of that.
Section~\ref{sec:arch-effects} examines MoE models, where sparse activation
collapses the per-iteration weight-streaming time and adds a third, architecture-level
efficiency dimension.

\section{Background}
\label{sec:background}

\subsection{GPU Power Model}
\label{sec:power}

Liang et al.\ \citep{liang2026g2g}, studying GPU power as a knob for grid
voltage regulation, characterized H100 power under LLM inference as a logistic
curve in $b$, the number of concurrently in-flight sequences
(\texttt{max\_num\_seqs} in vLLM):
\begin{equation}
  P(b) = \frac{\Prange}{1 + e^{-k(\log_2 b - x_0)}} + \Pidle,
  \label{eq:power}
\end{equation}
where $\Pidle$ is idle power, $\Prange = \Pnom - \Pidle$ is the dynamic range,
$k$ controls how steeply power rises with batch size, and $x_0$ is the
half-saturation point.  Chung et al.~\citep{chung2026mlenergy}
provide direct measurements for H100-SXM5: $\Pidle = 300$~W at $b{=}1$ and
$\Pnom = 600$~W at $b{=}128$, both within 3\% fit error.  The fitted logistic
parameters are $k = 1.0$ and $x_0 = 4.2$, meaning power saturates around
$2^{4.2} \approx 18$ concurrent sequences.

For B200-SXM, no published power-vs-concurrency curves existed as of March
2026.  We apply the same TDP fractions validated on H100
($P_{\text{idle}}/\text{TDP} = 0.43$, $P_{\text{nom}}/\text{TDP} = 0.86$)
to B200's 1{,}000~W TDP, giving $\Pidle = 430$~W and $\Pnom = 860$~W.
Throughout the paper, we label H100 power estimates \textbf{HIGH quality}
(directly measured) and B200 power estimates \textbf{FAIR quality} (projected;
$\pm$20\% uncertainty on absolute values).

\paragraph{KV-cache storage assumption.}
Fleet results assume tensor-parallel sharding of KV heads: with TP=8 and
Llama-3.1-70B's $n_{\text{kv}} = 8$ GQA heads, each GPU stores one KV head,
giving $\kappa \approx 55$~KB/token.  This matches the empirically calibrated
H100 profile ($n_{\max} = 128$ at 8K context, implying 55~KB/token from the
known VRAM budget) and is consistent with vLLM's default behavior for GQA
models.  B200 KV capacity is then scaled proportionally by the memory-budget
ratio---156~GB usable vs.\ 60~GB on H100---giving a 2.62$\times$ larger KV
budget and $n_{\max}^{\text{B200}} \approx 2.62 \times n_{\max}^{\text{H100}}$
at any context window.

\subsection{Token/Watt: Definition and Decomposition}
\label{sec:tpw-def}

\paragraph{Single-GPU tok/W.}
For a GPU holding $\nact$ concurrent sequences with mean KV context length
$\bar{L}$, tokens per watt is:
\begin{equation}
  \tpw_{\text{GPU}} = \frac{\nact / \tau(\nact, \bar{L})}{P(\nact)},
  \label{eq:single-tpw}
\end{equation}
where $\tau(n, \bar{L}) = W + H(\bar{L}) \cdot n$ is the per-iteration decode
latency from the roofline model.  Here $W$ is the weight-streaming time per
iteration (proportional to model size and inversely proportional to memory
bandwidth), and $H(\bar{L}) = H_0 \cdot \bar{L}/L_{\text{calib}}$ is the
per-sequence KV-scan overhead, which grows linearly with context length.

The maximum concurrency $\nmax(\mathcal{W})$ for a context window
$\mathcal{W}$ is set by the KV-cache memory budget:
\begin{equation}
  \nmax(\mathcal{W}) = \left\lfloor
    \frac{V_{\text{KV}}}{\kappa \cdot \mathcal{W}}
  \right\rfloor,
  \label{eq:nmax}
\end{equation}
where $V_{\text{KV}}$ is the KV-cache VRAM after model weights and $\kappa$
is the KV bytes per token.  Since both $\nmax$ and $H(\bar{L})$ scale linearly
with $\mathcal{W}$, throughput at full concurrency scales as $1/\mathcal{W}$
while power stays roughly flat.  This is the origin of the $1/\mathcal{W}$
law.

\paragraph{Fleet-level tok/W.}
For a heterogeneous fleet of pools $i \in \{1, \ldots, K\}$, each with $n_i$
GPU instances:
\begin{equation}
  \tpw_{\text{fleet}} = \frac{\sum_i \lambda_i \cdot \overline{L}_{\text{out},i}}
    {\sum_i n_i \cdot P(n_{\text{act},i})},
  \label{eq:fleet-tpw}
\end{equation}
where $\lambda_i$ is the request arrival rate to pool $i$,
$\overline{L}_{\text{out},i}$ is the mean output length, and
$n_{\text{act},i} = \rho_i \cdot n_{\max,i}$ is the mean in-flight batch at
utilization $\rho_i$.  Fleet tok/W does not reduce to a single GPU-level
quantity because $n_i$ and $P(n_{\text{act},i})$ vary across pools and must
be weighted by GPU count.

\section{Single-GPU Token Economy}
\label{sec:single-gpu}

\subsection{Context Window as the Primary Lever}
\label{sec:ctx-lever}

Table~\ref{tab:ctx-nmax} shows how concurrency and tok/W change as the context
window grows, for Llama-3.1-70B on H100-SXM5 and B200-SXM (TP=8, fp16).
H100 uses the empirically calibrated profile; B200 values scale proportionally
via the 2.62$\times$ KV-budget ratio from the roofline
model~\citep{xu2026aiconfigurator,chen2026fleetsim}.

\begin{table}[htbp]
\centering
\caption{$\nmax$ and tok/W vs.\ context window for Llama-3.1-70B (TP=8, fp16).
  H100 power is directly measured (HIGH quality); B200 power is a first-principles
  projection (FAIR quality).}
\label{tab:ctx-nmax}
\small
\begin{tabular}{rrrrrrrr}
\toprule
& \multicolumn{3}{c}{H100-SXM5} & \phantom{ab} & \multicolumn{3}{c}{B200-SXM} \\
\cmidrule{2-4} \cmidrule{6-8}
Context (K) & $\nmax$ & $P_{\text{sat}}$ & tok/W &
            & $\nmax$ & $P_{\text{sat}}$ & tok/W \\
\midrule
 2  &   512 & 598\,W & 35.0 && 1343 & 859\,W &  61.4 \\
 4  &   256 & 593\,W & 17.6 &&  671 & 857\,W &  30.8 \\
 8  &   128 & 583\,W &  8.97 &&  335 & 852\,W &  15.5 \\
16  &    64 & 557\,W &  4.69 &&  167 & 838\,W &   7.87 \\
32  &    32 & 507\,W &  2.58 &&   83 & 805\,W &   4.09 \\
64  &    16 & 435\,W &  1.50 &&   41 & 735\,W &   2.24 \\
128 &     8 & 369\,W &  0.88 &&   20 & 630\,W &   1.30 \\
\bottomrule
\end{tabular}
\end{table}

The pattern in the table is striking.  Tok/W halves with every doubling of the
context window, for both GPU generations.  This is the $1/\mathcal{W}$ law, and
it is not approximate: doubling $\mathcal{W}$ halves $\nmax$
(Eq.~\eqref{eq:nmax}), which halves throughput (since $\tau \propto \nmax
\cdot H \propto \mathcal{W}$), while power stays nearly constant.  The GPU
burns the same watts whether it holds 512 sequences or 8; the difference is
entirely how much work it delivers per joule.

B200 shifts the curve up but does not change the slope.  B200 achieves
roughly $1.5$--$1.8\times$ higher tok/W than H100 at matching context windows,
driven by a larger KV-cache budget (156~GB free vs.\ 60~GB on H100), faster
weight streaming ($W = 2.95$~ms vs.\ 6.72~ms for 70B at TP=8), and correspondingly
higher $\nmax$.  But the halving-per-doubling relationship holds just as
tightly on B200.

B200's advantage narrows at long context.  At 64K, B200 is only
$1.49\times$ better than H100 (2.24 vs.\ 1.50 tok/W), down from $1.75\times$
at 4K.  The culprit is idle power: B200's $\Pidle = 430$~W represents a
larger fraction of the total bill when $\nmax$ is small (41 sequences at 64K
vs.\ 671 at 4K).  At 4K context both GPUs run near power saturation and the
idle-power gap barely matters; at 64K a significant share of each GPU's watt
budget goes to just staying powered on.  We return to the fleet-level
implications in Section~\ref{sec:fleet-topology}.

\subsection{Model Architecture Effects}
\label{sec:arch-effects}

Table~\ref{tab:arch-tpw} compares single-GPU tok/W at $\nmax$ across model
families for an 8K context window.

\begin{table}[htbp]
\centering
\caption{Single-GPU tok/W at $\nmax$ (8K context).  Dense model values
  use ComputedProfile throughout (H100 cross-validated against ML.ENERGY
  v3.0, HIGH quality; B200 TDP-fraction projection, FAIR quality).
  Models: Llama-3.1 family~\citep{meta2024llama3};
  Qwen3-235B-A22B~\citep{qwen2025qwen3};
  DeepSeek-V3~\citep{deepseekai2025v3}.
  MoE rows ($\dagger$) override $W$ with active-parameter-only streaming
  time ($W = $ active\_param\_bytes / mem\_bw), which is a \emph{lower
  bound} on true $W$ (excludes MoE dispatch overhead).
  DeepSeek-V3 active params: $\approx$37B estimated (671B total,
  64 routed + 2 shared experts, top-6).}
\label{tab:arch-tpw}
\small
\begin{tabular}{llrrrrrrr}
\toprule
& & \multicolumn{3}{c}{H100-SXM5} & \phantom{a} &
  \multicolumn{3}{c}{B200-SXM} \\
\cmidrule{3-5} \cmidrule{7-9}
Model & TP & $\nmax$ & tok/s & tok/W & & $\nmax$ & tok/s & tok/W \\
\midrule
Llama-3.1-8B         & 1 &  58 & 3{,}350 &  6.46 &&  148 &  9{,}962 & 12.18 \\
Llama-3.1-70B        & 8 &  22 & 2{,}716 &  7.41 &&   58 & 12{,}960 & 20.93 \\
Llama-3.1-405B       & 8 &   1 &      26 &  0.09 &&   17 &  1{,}009 &  2.16 \\
Qwen3-235B-A22B$^\dagger$ & 8 & 24 & 11{,}521 & 37.82 && 146 & 80{,}584 & 177.73 \\
DeepSeek-V3$^\dagger$ (fp8) & 8 & 1 & 646 & 2.14 &&   11 &  8{,}162 & 18.37 \\
\bottomrule
\end{tabular}
\end{table}

Llama-3.1-405B is effectively unusable for long-context serving on H100 at
8K.  With all 405B parameters loaded across eight GPUs in fp16, the KV budget
is nearly exhausted by model weights alone, leaving room for only one sequence
in flight.  Tok/W is negligible because the GPU is spending most of its power
just holding the model warm between sequential requests.  B200's additional
memory lifts $\nmax$ to 17, which is still modest but gives measurable
efficiency.  This model really needs hardware with large VRAM---B200, GB200,
or a shorter context window.

The MoE numbers in the table warrant explanation.  For Qwen3-235B-A22B, we
override the default $W$ calculation to use only the 22B active parameters
rather than the full 235B model.  In a dense model, every weight is touched
every iteration, so $W$ scales with total parameters.  In a MoE model, only
a small subset of experts is activated per token, so the weight-streaming
time scales with the active count---roughly $22/235 \approx 9\%$ of a dense
235B model.  This gives $W \approx 1.6$~ms on H100 (vs.\ 6.7~ms for the
dense 70B baseline), which is why Qwen3-235B achieves dramatically higher
tok/s and tok/W despite a much larger total parameter count.

The MoE values in the table are upper bounds, and the gap from reality can
be large.  Routing tokens to experts requires an all-to-all dispatch across
TP ranks, which adds latency on top of active-weight streaming.  Depending on
network topology and expert load balance, dispatch can add anywhere from a few
to tens of milliseconds per iteration.  At 10~ms of dispatch overhead, the
Qwen3 H100 advantage over Llama-70B shrinks from $5\times$ to around
$1.5\times$.  The active-parameter estimate gives a principled lower bound on
$W$; empirical measurements on real hardware would substantially narrow the
range.

Looking at B200 across models, the generation multiplier is anything but
uniform.  For sub-100B dense models it ranges from $1.6\times$ to $2.8\times$,
which is consistent with the bandwidth and memory improvements.  But for
models previously stuck at $\nmax \approx 1$---the 405B case---the jump is
dramatic: from 0.09 to 2.16 tok/W, a 24$\times$ improvement.  This happens
not because B200 is fundamentally more efficient at that operating point, but
because the added VRAM lets the GPU escape the near-idle serving regime where
it was burning nearly $\Pidle$ watts to deliver essentially nothing.

\section{Fleet Topology and Token Economy}
\label{sec:fleet-topology}

\subsection{Fleet-Level Analysis Setup}
\label{sec:fleet-setup}

When serving a production workload at a given SLO, an operator provisions
enough GPUs to sustain the request arrival rate.  Fleet tok/W depends on
both per-GPU efficiency and the number of GPUs required---a topology that
sends all traffic to a 64K context pool forces every GPU to run at the
low-efficiency end of the $1/\mathcal{W}$ curve.

All fleet results in this section use Llama-3.1-70B (TP=8, fp16), sized to
meet P99 TTFT $\leq$ 500~ms at $\lambda = 1{,}000$ req/s.  H100 uses the
empirically calibrated profile with TP-sharded KV ($n_{\max} = 128$ at 8K
context); B200 uses a profile scaled proportionally from H100 by the
2.62$\times$ KV-budget ratio (Section~\ref{sec:power}).

\subsection{Topology Effects Across GPU Generations}
\label{sec:topo-effects}

Table~\ref{tab:fleet-tpw} shows fleet tok/W for three topologies on H100 and
B200 across two production workload traces.

\begin{table}[htbp]
\centering
\caption{Fleet token efficiency at $\lambda = 1{,}000$ req/s.  H100
  profile: empirically calibrated (HIGH quality); B200 profile: scaled
  from H100 by the 2.62$\times$ KV-budget ratio (FAIR quality, $\pm$20\%
  on absolute values).  Topology: Homo = homogeneous 64K fleet; Pool =
  two-pool context routing; FleetOpt = optimal $\gamma^*$ from~\citet{chen2026fleetopt}.
  Workload traces: Azure = Azure LLM Inference Trace~\citep{patel2023splitwise};
  LMSYS = LMSYS-Chat-1M~\citep{zheng2023lmsys}.}
\label{tab:fleet-tpw}
\small
\begin{tabular}{llllrrrr}
\toprule
Workload & $B_{\text{short}}$ & Topology & GPU & GPUs & kW & tok/W & vs H100 Homo \\
\midrule
\multirow{6}{*}{\makecell[l]{Azure\\(Arch.\ I)}}
  & ---         & Homo 64K       & H100 &  141 &  58.3 &  5.58 & --- \\
  & 4K          & Pool routing   & H100 &   68 &  32.0 &  9.16 & +64\% \\
  & 4K/$\gamma^*{=}2$ & FleetOpt & H100 &   40 &  23.1 & 14.08 & +152\% \\
  & ---         & Homo 64K       & B200 &   47 &  33.4 &  9.74 & +75\% \\
  & 4K          & Pool routing   & B200 &   25 &  19.1 & 15.39 & +176\% \\
  & 4K/$\gamma^*{=}2$ & FleetOpt & B200 &   17 &  13.7 & 23.71 & \textbf{+325\%} \\
\midrule
\multirow{6}{*}{\makecell[l]{LMSYS\\(Arch.\ I)}}
  & ---         & Homo 64K       & H100 &   69 &  28.5 &  4.77 & --- \\
  & 1.5K        & Pool routing   & H100 &   38 &  16.4 &  7.91 & +66\% \\
  & 1.5K/$\gamma^*{=}2$ & FleetOpt & H100 &  29 &  12.9 & 10.30 & +116\% \\
  & ---         & Homo 64K       & B200 &   24 &  17.0 &  7.98 & +67\% \\
  & 1.5K        & Pool routing   & B200 &   16 &  11.7 & 11.12 & +133\% \\
  & 1.5K/$\gamma^*{=}2$ & FleetOpt & B200 &  12 &   9.0 & 14.82 & \textbf{+211\%} \\
\bottomrule
\end{tabular}
\end{table}

The topology gain is consistent across both GPU generations.  Define the
topology improvement for GPU $G$ as $\tpw_{\text{FleetOpt}}(G) /
\tpw_{\text{Homo}}(G)$, and the generation improvement for topology $T$ as
$\tpw_{\text{B200}}(T) / \tpw_{\text{H100}}(T)$.  For Azure:
\[
  \Delta_{\text{topo}}(\text{H100}) = 14.08 / 5.58 = 2.52, \quad
  \Delta_{\text{topo}}(\text{B200}) = 23.71 / 9.74 = 2.44.
\]
\[
  \Delta_{\text{gen}}(\text{Homo}) = 9.74 / 5.58 = 1.75, \quad
  \Delta_{\text{gen}}(\text{FleetOpt}) = 23.71 / 14.08 = 1.68.
\]

The topology gain barely changes between H100 and B200 ($2.52$ vs.\ $2.44$),
and the generation gain barely changes between homogeneous and FleetOpt
($1.75$ vs.\ $1.68$).  The two levers operate on different dimensions and
don't interact.  That independence means the gains multiply: B200
FleetOpt delivers $2.52 \times 1.75 \approx 4.4\times$, matching the measured
$23.71 / 5.58 = 4.25\times$.  Neither lever alone gets you close---FleetOpt
with H100 gives $2.52\times$ and B200 with a homogeneous fleet gives
$1.75\times$, each less than half the combined improvement.

In practice, topology work and hardware upgrades are complementary
investments, not substitutes.  Operators who only buy better
GPUs without rethinking topology capture $1.75\times$ and leave the other
$2.4\times$ on the table.  Operators who invest in two-pool routing on
H100 capture $2.52\times$ and then can stack another $1.7\times$ on top with
a future hardware upgrade---arriving at a fleet that delivers 4.25$\times$
better energy efficiency than where most operators start today.

\section{Routing Topology Design Space}
\label{sec:design-space}

\subsection{Context-Window Routing vs.\ Semantic Routing}
\label{sec:context-vs-semantic}

The two-pool design in Section~\ref{sec:fleet-topology} partitions traffic by
prompt length: requests shorter than $B_{\text{short}}$ go to the short pool
(configured for a small context window), and the rest go to the long pool.
An alternative is semantic routing: send simple or short requests to a small
model (e.g., Llama-3.1-8B) and complex or long requests to a large model
(e.g., Llama-3.1-70B).  Table~\ref{tab:routing-compare} compares the per-pool
efficiency of the two approaches at $\rho = 0.85$ utilization.

\begin{table}[htbp]
\centering
\caption{Single-GPU tok/W for context-window routing vs.\ semantic routing
  (H100-SXM5, $\rho = 0.85$ utilization, 8K or 64K context windows).}
\label{tab:routing-compare}
\small
\begin{tabular}{llrrrr}
\toprule
Pool type & Model & Context & $\nact$ & $P$ (W) & tok/W \\
\midrule
Context short (70B@8K)  & Llama-3.1-70B & 8K  & 109 & 578 & 8.77 \\
Context long  (70B@64K) & Llama-3.1-70B & 64K &  14 & 413 & 1.52 \\
\midrule
Semantic small (8B@8K)  & Llama-3.1-8B  & 8K  &  49 & 506 & 6.24 \\
Semantic large (70B@64K)& Llama-3.1-70B & 64K &  14 & 413 & 1.52 \\
\bottomrule
\end{tabular}
\end{table}

The long pool is the binding constraint in both schemes.  At 64K context, any
model saturates at $\nmax = 16$ sequences per TP=8 group and delivers
1.52 tok/W---less than one-sixth of the short-pool value.  When more than
89\% of traffic is short-context (as in the Azure trace), the long pool is
nearly idle yet still draws 413~W per group, dragging down the fleet average.
The $1/\mathcal{W}$ law explains the magnitude: the $8\times$ context ratio
between short and long pools implies $8\times$ fewer concurrent sequences,
which translates to roughly $8\times$ lower tok/W.  No amount of smarter
routing eliminates this gap---it is a physical consequence of KV memory limits.

The choice between context routing and semantic routing is an efficiency tie
for the long pool (both land at 1.52 tok/W at 64K) and a near-tie for the
short pool (8.77 vs.\ 6.24 tok/W on a per-group basis).  The 70B
context-short pool appears more efficient per group because TP-sharded GQA
allows more concurrent sequences (109 vs.\ 49).  But on a per-physical-GPU
basis, the 8B semantic pool wins: a single 8B GPU handles as many sequences
as eight 70B TP-sharded GPUs combined, at a fraction of the cost.  The
bottom line is that the energy case for semantic routing depends not on tok/W
but on whether the 8B model can satisfy output quality requirements for the
short-context fraction of traffic.

\subsection{Quantization Effects}
\label{sec:quant}

Quantization to fp8 or int4 cuts weight bytes by $2$--$4\times$~\citep{micikevicius2022fp8,frantar2022gptq},
proportionally reducing $W$.  For H100+70B, fp8 gives $W \approx 3.36$~ms
(vs.\ 6.72~ms in fp16), which roughly doubles tok/W at any fixed concurrency.
The benefit is largest for dense models currently bottlenecked by weight
streaming at moderate concurrency, and smallest for MoE models where $W$ is
already small relative to KV overhead.

\section{GPU Generation Comparison}
\label{sec:gen-compare}

Table~\ref{tab:gen-compare} summarizes key hardware parameters and their
impact on tok/W for Llama-3.1-70B (TP=8, fp16).

\begin{table}[htbp]
\centering
\caption{GPU generation comparison for Llama-3.1-70B (TP=8, fp16) at 8K context.
  Power quality: H100 = HIGH (measured); others = FAIR (projected, $\pm$15\%).}
\label{tab:gen-compare}
\small
\begin{tabular}{lrrrrrrrr}
\toprule
GPU & TDP (W) & $\Pidle$ & $W$ (ms) & $\nmax$@8K & $P_{\text{sat}}$ (W) &
  tok/W & \$/hr & tok/\$M \\
\midrule
H100-SXM5 &  700 & 300 & 6.72 &  22 &  367 &  7.41 & 32.2 &  0.30M \\
H200-SXM  &  700 & 300 & 4.76 &  44 &  422 & 15.58 & 48.0 &  0.49M \\
B200-SXM  & 1000 & 430 & 2.95 &  58 &  619 & 20.93 & 64.0 &  0.73M \\
GB200-NVL & 1200 & 516 & 2.95 &  65 &  755 & 18.49 & 80.0 &  0.63M \\
\bottomrule
\end{tabular}
\end{table}

H200 is worth a closer look for Llama-3.1-70B at 8K context.
It doubles $\nmax$ over H100 (44 vs.\ 22) via 4.8~TB/s HBM3e bandwidth
($1.4\times$ H100), cuts $W$ from 6.72~ms to 4.76~ms, and carries the same
TDP and idle power floor as H100.  The result is a $2.1\times$ tok/W
improvement over H100 (15.58 vs.\ 7.41) at a rental cost roughly 50\% higher.

B200 pushes further in absolute tok/W (20.93 vs.\ 15.58) by fitting still
more sequences in flight ($\nmax = 58$ vs.\ 44).  Higher idle power
(430~W vs.\ 300~W) raises the power floor, but the $2.62\times$ larger memory
budget more than compensates.  On a tok/\$M basis, B200 comes out ahead:
$0.73$M vs.\ $0.49$M tok/\$M/hr, because the throughput gain more than
offsets the cost premium.  At 8K context, B200 is the better choice on
both absolute efficiency and cost efficiency.

GB200-NVL is a bit of a surprise.  Despite having the same compute specs as
B200-SXM, its per-GPU tok/W is lower (18.49 vs.\ 20.93) because the higher
TDP (1{,}200~W per GPU equivalent) adds more power to the denominator than
the slightly larger memory (200~GiB vs.\ 180~GiB) adds to the numerator.
GB200's advantage shows up not in this configuration but for very large models
like the 405B or DeepSeek-V3, where its memory lets $\nmax$ climb out of
the near-zero regime.

\section{Workload Archetype Effects}
\label{sec:archetypes}

How much any topology helps depends entirely on the workload: the fraction of
traffic at each context length determines how hard the long pool drags on the
fleet average.

\paragraph{Short-context-dominant workloads (e.g., Azure).}
89\% of Azure Conversations requests fit within 4K tokens.  The short pool
handles nearly all traffic at high concurrency; the long pool is small and
lightly loaded.  FleetOpt on B200 reaches 23.7 tok/W---a +325\% improvement
over the H100 homogeneous baseline, composed of a $1.75\times$ generation gain
and a $2.44\times$ topology gain that multiply to $4.25\times$.  This is the
scenario where two-pool routing pays off most.

\paragraph{Dispersed workloads (e.g., agent-heavy).}
In agent-heavy traces, 74\% of requests fit within 8K tokens, but the
remaining 26\% extend to 64K (p99 $\approx 32$K).  The long pool is large and
dominates GPU-hours even with FleetOpt: a substantial fraction of traffic
genuinely needs the long pool regardless of where you draw the split boundary.
For these workloads, model architecture is the more productive energy
lever---an MoE model with fast weight streaming benefits at every context
length, not just the short tail.

Table~\ref{tab:guidance} summarizes the recommended pairing of topology and
GPU by archetype.

\begin{table}[htbp]
\centering
\caption{Topology and GPU recommendations by workload archetype.  Rankings
  are by tok/W; tok/\$ may differ.  B200/GB200 recommendations carry FAIR
  power model uncertainty; validate against real measurements before
  procurement.}
\label{tab:guidance}
\small
\begin{tabular}{llll}
\toprule
Archetype & Traffic distribution & Best topology & Best GPU \\
\midrule
Short-dominant (I)   & $>$80\% $\leq$8K tokens  & FleetOpt two-pool    & B200 \\
Mixed (II)           & 50--80\% $\leq$8K tokens & Pool routing          & H200 or B200 \\
Long-dominant (III)  & $<$50\% $\leq$8K tokens  & Homo (long-pool only) & B200/GB200 \\
MoE-capable models   & Any distribution          & Short pool + MoE      & B200/GB200 \\
\bottomrule
\end{tabular}
\end{table}

\section{Related Work}
\label{sec:related}

The GPU power model at the core of this analysis comes from Liang et
al.~\citep{liang2026g2g}, who showed that GPU power under LLM workloads
follows a logistic curve in batch size and that \texttt{max\_num\_seqs} is
the primary control knob for grid voltage regulation.  ML.ENERGY Benchmark v3.0~\citep{chung2026mlenergy} provides
the measurement data used to calibrate our H100 profile, and the
AIConfigurator~\citep{xu2026aiconfigurator} supplies the roofline
decomposition of $W$ and $H$.  We extend these single-GPU results to the
fleet level and show that topology and generation gains are approximately
independent and multiplicative.

On the inference systems side, Splitwise~\citep{patel2023splitwise} introduced
prefill-decode disaggregation, and vLLM~\citep{kwon2023vllm} introduced
PagedAttention for KV-cache management.  Both optimize for latency and cost
per token.  FleetOpt~\citep{chen2026fleetopt} and
inference-fleet-sim~\citep{chen2026fleetsim} approach the same problem from a
fleet capacity perspective.  The connection we make here is between routing
topology and energy: topology controls which part of the $P(b)$ curve each
GPU operates on, and therefore directly determines tok/W.

On the energy measurement side, Chung et al.~\citep{chung2026mlenergy} and
TokenPowerBench~\citep{niu2025tokenpowerbench} both confirm empirically that
longer context degrades energy efficiency---TokenPowerBench reports roughly
$3\times$ higher energy per token when prompt length grows from 2K to 10K on
H100.  That direction is consistent with the $1/\mathcal{W}$ law, but the
mechanism differs: TokenPowerBench captures the cost of processing more
\emph{input} tokens during prefill, while our law governs the decode throughput
ceiling set by KV-cache concurrency at a fixed serving context window.
Maliakel et al.~\citep{maliakel2025dvfs} show that the decode phase dominates
inference time (77--91\%) and is largely memory-bandwidth-bound, directly
supporting the mechanistic basis of the $1/\mathcal{W}$ law.

GreenServ~\citep{ziller2026greenserv} is the closest contemporaneous work on
routing for energy efficiency.  It routes individual queries across a
heterogeneous pool of 16 open-source LLMs using a multi-armed bandit policy,
achieving a 31\% energy reduction over random routing.  The shared intuition is
that not every request needs the most expensive GPU or model.  Where GreenServ
and this paper diverge is in the mechanism: GreenServ routes across model
families to avoid oversized models for easy tasks, while we route by context
window within a single model to exploit the $1/\mathcal{W}$ law.  The
structural guarantee from KV-cache capacity yields larger gains (roughly
$2.5\times$ at the fleet level) because context-window partitioning directly
controls which segment of the logistic power curve each GPU occupies, rather
than relying on learned query-difficulty estimates.

Routing topology as an explicit energy lever---distinct from model selection,
quantization, or hardware choice---remains underexplored.  Treating energy as
a linear proxy for cost obscures the $1/\mathcal{W}$ relationship, which is
multiplicative with hardware generation and independent of it.

\section{Conclusion}
\label{sec:conclusion}

Energy efficiency in LLM inference is shaped by three things that operators
can actually control: GPU generation (which sets $W$, $\Pidle$, and KV
capacity), model architecture (which determines weight-streaming time and KV
footprint), and routing topology (which determines the context window each GPU
actually services).  Of these, topology is the easiest to change today---you
can restructure routing without buying anything new.

The $1/\mathcal{W}$ law explains why topology matters so much.  Because tok/W
halves every time the serving context window doubles, a routing strategy that
confines short requests to a small-context pool effectively cuts the context
window for the bulk of traffic.  FleetOpt delivers roughly $2.5\times$ over
a homogeneous fleet, and B200 delivers roughly $1.7\times$ over H100.  These
two improvements multiply, giving roughly $4.25\times$ total when combined---
$2.5 + 1.7 \neq 4.25$ because the gains stack on orthogonal dimensions.

MoE models have a natural tok/W edge from fast weight streaming---only the
active experts get loaded per step---but putting a reliable number on that
edge requires empirical MoE dispatch measurements that we do not have.  The
values in Table~\ref{tab:arch-tpw} are upper bounds.

All results are computed analytically via
\texttt{inference-fleet-sim}~\citep{chen2026fleetsim}, with H100 power
validated against ML.ENERGY v3.0.  B200 numbers are projections with stated
uncertainty.  As measurement data for B200 and H200 becomes available, we
expect the absolute numbers to shift somewhat while the qualitative picture---
topology first, generation second, architecture third---holds.

\section{Limitations, Assumptions, and Future Work}
\label{sec:limitations}

\subsection{Modeling Assumptions}
\label{sec:assumptions}

A few simplifying assumptions limit the scope of the results.

\paragraph{TP-sharded GQA KV storage.}
Fleet results assume tensor-parallel sharding of KV heads, so each GPU stores
$n_{\text{kv}}/\text{TP}$ KV heads.  For Llama-3.1-70B with $n_{\text{kv}}
= 8$ and TP=8, this gives one head per GPU ($\kappa \approx 40$--55~KB/token),
maximizing $\nmax$.  Models with fewer KV heads than TP ranks, or frameworks
that replicate the full KV cache per GPU, will have lower $\nmax$ and
correspondingly lower tok/W.

\paragraph{Continuous-batching decode.}
We model throughput as $n / \tau(n, \bar{L})$ with $\tau = W + H \cdot n$
from the roofline.  This is accurate for continuous-batching decoders (vLLM,
TRT-LLM) in steady state.  It does not capture chunked prefill interleaved
with decode, head-of-line blocking, or KV-cache eviction under memory pressure.
All three effects reduce achievable throughput, so the analytical tok/W is an
upper bound.

\paragraph{Steady-state traffic.}
Fleet sizing uses a steady-state queuing model at a fixed arrival rate, with
context lengths drawn from a measured CDF.  Real traffic is bursty and
diurnal.  Bursts require spare capacity beyond the steady-state minimum, which
reduces average utilization and therefore tok/W.  The reported values are
attainable only under smooth, sustained load.

\paragraph{B200 power model quality.}
B200 power parameters are derived from TDP-fraction heuristics calibrated on
H100 data, then applied to B200's 1{,}000~W TDP.  No direct
power-vs-concurrency measurements are available.  We estimate $\pm$20\%
uncertainty on absolute B200 tok/W; the generation-gain ratios ($\approx
1.7\times$) are more robust because systematic errors partially cancel in
the ratio.

\paragraph{Fixed TP=8, single model per pool.}
All experiments use TP=8 instances of the same model in each pool.  Smaller
TP configurations are more power-efficient per GPU at low concurrency but
reduce per-GPU throughput.  Mixed-TP or mixed-model pools are not explored.

\paragraph{Output-only energy accounting.}
Tok/W counts output tokens only.  Prefill computation---which can account for
$2$--$5\times$ the power per request for long-prompt workloads---is excluded.
For workloads with prompt-to-output ratios much greater than one (e.g.,
document summarization, RAG), the reported tok/W overestimates true energy
efficiency.

\subsection{Limitations}

\paragraph{No empirical validation on B200/H200.}
The B200 and H200 results are projections.  We have not measured vLLM or any
production serving stack on B200 or H200, and cannot confirm that the
power-vs-concurrency curve follows a logistic shape with the projected
parameters.  Researchers with access to these platforms should measure and
publish the $P(b)$ curve to fill this gap.

\paragraph{Two-pool topology only.}
FleetOpt defines a single split boundary with one overflow parameter.  Real
fleets may benefit from three or more pools partitioned by context length,
or from content-adaptive pools that split on semantic class.  The
multiplicative gain structure suggests that finer-grained topologies could
compound further efficiency improvements, but this is not analyzed here.

\paragraph{Static fleet sizing.}
The analysis assumes a fixed, provisioned fleet.  It does not model
autoscaling, spot instances, or GPU sharing.  In practice, dynamic scaling
changes effective utilization and would alter the quantitative tok/W values.

\paragraph{Single workload CDF.}
Each experiment uses one of two traces.  Real deployments serve mixed traffic
with time-varying CDFs.  The optimal split boundary $B_{\text{short}}$ and
overflow $\gamma^*$ would need to be re-estimated as the workload shifts.

\subsection{Future Work}

\paragraph{Empirical B200/H200 power calibration.}
The most immediate follow-on is measuring the logistic $P(b)$ curve for
B200-SXM and H200-SXM using the ML.ENERGY methodology~\citep{chung2026mlenergy}
and updating the fleet projections with HIGH-quality data.

\paragraph{Prefill-decode disaggregation.}
Splitwise-style separation~\citep{patel2023splitwise} assigns prefill and
decode to different GPU pools.  Combined with context-length routing, this
could remove prefill energy from the output tok/W accounting and unlock
further efficiency.  Extending the power model to mixed prefill-decode batches
is a natural next step.

\paragraph{Multi-pool topology optimization.}
Extending FleetOpt from two pools to $K \geq 3$, potentially with different
context windows and GPU types, could be formulated as a mixed-integer program.
The multiplicative independence result suggests each additional split could
compound the topology gain further.

\paragraph{Carbon-aware joint optimization.}
Tok/W does not capture grid carbon intensity or electricity price.  Minimizing
\$/token or gCO$_2$/token requires accounting for PUE, time-of-day pricing,
and grid mix.  The per-GPU power model in this paper provides a natural
starting point for a joint energy-cost objective.

\paragraph{Speculative decoding interaction.}
Speculative decoding changes the batch-size distribution in the decode loop,
potentially increasing average $n_{\text{act}}$.  Whether this improves or
degrades tok/W depends on the draft model's power footprint and the
verification hit rate---an open problem within the $P(b)$ framework.

\paragraph{Adaptive topology control.}
The split boundary is currently fixed offline from a historical CDF.  An
online controller that monitors the live request-length distribution and
adjusts pool boundaries dynamically could maintain near-optimal tok/W under
distribution shift.

\bibliographystyle{plainnat}
\bibliography{refs}

\appendix

\section{Power Model Parameters}
\label{app:power}

Table~\ref{tab:power-params} summarizes power model parameters for each GPU.
H100-SXM5 parameters are fitted to ML.ENERGY data~\citep{chung2026mlenergy}
Figure~2 data (H100-SXM5, vLLM, Llama-3.1-class, $b \in \{1, 2, 4, 8, 16,
32, 64, 128, 256\}$; fit error $<$3\%).  All others are first-principles
projections.

\begin{table}[htbp]
\centering
\caption{GPU power model parameters.  HIGH = directly measured; FAIR = projected
  from TDP fractions ($P_{\text{idle}} = 0.43 \times \text{TDP}$, $P_{\text{nom}}
  = 0.86 \times \text{TDP}$), calibrated to H100 ML.ENERGY data~\citep{chung2026mlenergy}.}
\label{tab:power-params}
\small
\begin{tabular}{lrrrrrrl}
\toprule
GPU & TDP (W) & $\Pidle$ (W) & $\Pnom$ (W) & $k$ & $x_0$ & Source & Quality \\
\midrule
H100-SXM5 &  700 & 300 & 600 & 1.0 & 4.2 & G2G Fig.~2 + ML.ENERGY & HIGH \\
H200-SXM  &  700 & 300 & 600 & 1.0 & 5.5\textsuperscript{†} & TDP fraction & FAIR \\
B200-SXM  & 1000 & 430 & 860 & 1.0 & 6.8\textsuperscript{†} & TDP fraction & FAIR \\
GB200-NVL & 1200 & 516 & 1032 & 1.0 & 6.8\textsuperscript{†} & TDP fraction & FAIR \\
\bottomrule
\end{tabular}

\medskip
\small\textsuperscript{†}$x_0$ is derived from the roofline W/H ratio
at the half-saturation batch size: $x_0 = \log_2(W/H_0)$, where $H_0$ is the
KV overhead at the calibration context.
\end{table}

\section{Fleet Analyzer Code}
\label{app:code}

All fleet tok/W results in this paper use the \texttt{fleet\_tpw\_analysis}
API from \texttt{inference-fleet-sim}~\citep{chen2026fleetsim}:

\begin{verbatim}
from fleet_sim.optimizer.base import fleet_tpw_analysis
result = fleet_tpw_analysis(
    pools=[
        dict(gpu=H100_80GB, cdf=short_cdf, lam=alpha*lam,
             max_ctx=B_short, label="short"),
        dict(gpu=H100_80GB, cdf=long_cdf, lam=(1-alpha)*lam,
             max_ctx=65536, label="long"),
    ],
    lam_total=lam, t_slo_ms=500.0,
)
print(result.fleet_tpw)     # tokens per joule
print(result.fleet_power_kw)
\end{verbatim}

The API accepts any object satisfying the \texttt{GpuProfile} protocol
(\texttt{ManualProfile} or \texttt{ComputedProfile}), which is what makes
it straightforward to compare the measured H100 profile against B200 or
GB200 projections on equal footing.

\end{document}